\documentclass{revtex4}
\usepackage{amssymb}
\usepackage{amsfonts}
\usepackage{amsmath}

\input{tcilatex}

\begin{document}

\title{ Landau Quantization for An electric quadrupole moment of
Position-Dependent Mass Quantum Particles interacting with Electromagnetic
fields}
\author{ Zeinab Algadhi}
\email{zeinab.algadhi@emu.edu.tr}
\author{Omar Mustafa}
\email{omar.mustafa@emu.edu.tr}
\affiliation{Department of Physics, Eastern Mediterranean University, G. Magusa, north
Cyprus, Mersin 10 - Turkey,\\
Tel.: +90 392 6301378; fax: +90 3692 365 1604.}

\begin{abstract}
\textbf{Abstract:} Analogous to Landau quantization related to a neutral
particle possessing an electric quadrupole moment, we generalize such a
Landau quantization to include position-dependent mass (PDM) neutral
particles. Using cylindrical coordinates, the exact solvability of PDM
neutral particles with an electric quadrupole moment moving in
electromagnetic fields is reported. The interaction between the electric
quadrupole moment of a PDM neutral particle and a magnetic field in the
absence of an electric field is analyzed for two different radial
cylindrical PDM settings. Next, two particular cases of radial electric
fields $(\overrightarrow{E}=\frac{\lambda }{\rho }\widehat{\rho }\ and\ 
\overrightarrow{E}=\frac{\lambda \rho }{2}\widehat{\rho })$ are considered
to investigate their influence on the Landau quantization (of this system
using the same models of PDM settings). The exact eigenvalues and
eigenfunctions for each case are analytically obtained.


\end{abstract}

\maketitle

\section{\protect\bigskip Introduction}

The interaction of multipole moments with the electromagnetic fields has
attracted a lot of attention and produced fundamental quantum effects. For
example, the Aharonov-Bohm effect \cite{1,2,3,4} for a charged particle, the
scalar Aharonov-Bohm and He- Mckellar- Willens effects \cite{5,6,7,8,9,10,11}%
, for bound states \cite{12}, and Landau quantization \cite{13,14,15,16} for
an electric dipole moment of a neutral particle. Furthermore, recent studies
have investigated the interaction between the quadrupole moments of neutral
particle and external fields in several quantum systems such as geometric
quantum phases \cite{17}, noncommutative quantum mechanics \cite{19},
nuclear structure \cite{20,21}, atomic systems \cite{22,23,24,25,26,27},
molecular systems \cite{28,29,30}, and Landau quantization \cite{18,31,32,33}%
.

In particular, the study of the Landau quantization has recently been
applied to the quantum dynamics of an electric quadrupole moment \cite{18,31}%
. It investigated the possibility of achieving the Landau quantization for
neutral particles, resulting from the coupling of the electric quadrupole
moment with a magnetic field, making a similar minimal coupling with a
constant magnetic field \cite{1,2}. Moreover, they have discussed the
conditions necessary for the field configuration in order to achieve the
Landau quantization for neutral particles possessing an electric quadrupole
moment \cite{17,18}. It is shown \cite{18} that the field configuration in
the quadrupole system is dependent on the structure of the quadrupole tensor
( i.e., diagonal or non-diagonal), and has to be different in each case.
However, all the previous studies have used different methods for quantum
systems of multipole moments with constant mass. Such methods need to be
modified to include the spatial dependence of the mass.

On the other hand, quantum mechanical systems with position-dpendent mass
(PDM) have attracted attention over the years. Namely, the von Roos
Hamiltonian \cite{34} has been extensively investigated in the literature 
\cite{35,36,37,38,39,40,41,42,43,44,45,46,47,48,49,50,51,52,53,54,55}. Not
only because of its ordering ambiguity associated with the non-unique
representation of the kinetic energy operator, but also because of its
feasible applicability in many fields of physics. Recent studies on such PDM
charged particles in constant magnetic fields \cite{56,57,58,59,60}, and
position-dependent magnetic fields \cite{61} are carried out (using
different interaction potentials).\ To the best of our knowledge, however,
no studies have ever been considered to discuss the quantum mechanical
effects on PDM neutral particles possessing an electric quadrupole moment.
To fill this gap, we discuss in this work a quantum system that consists of
a PDM neutral particle with an electric quadrupole moment interacting with
external fields. We follow the discussion in \cite{18} and extend their idea
for PDM systems.

This paper is organized as follows. In section II, we start giving a brief
description of the quantum dynamics for a moving electric quadrupole moment
interacting with external fields with a constant mass as done in \cite{18},
and extend it to include the PDM case. In so doing, we use the very recent
result suggested by \cite{58,59} for the PDM-minimal-coupling and the
PDM-momentum operator. Furthermore, we discuss the possibility of achieving
the Landau quantization for such a system, and the separability of the
problem in the cylindrical coordinates $\left( \rho ,\phi ,z\right) $, under
azimuthal symmetrization, by considering that the field configurations and
the PDM settings are purely radial dependent as in \cite{54,55,59,60,61}. In
section III, we discuss a Landau levels analog for an electric quadrupole
moment interacting with an external magnetic field in the absence of
electric field. In the same section we obtain exact eigenfunctions and
eigenvalues for different PDM settings. We take into account, in section IV,
the effect of an electric field on the problem at hand, by choosing two
models for the radial electric field, a Coulomb-type electric field $%
\overrightarrow{E}=\frac{\lambda }{\rho }\widehat{\rho }$ and a linear-type
electric field $\overrightarrow{E}=\frac{\lambda \rho }{2}\widehat{\rho }$.
Finally, we report exact solutions of the radial Schr\"{o}dinger equation
for both case of an electric field and for the same PDM settings presented
in previous section. Our conclusion is given in section V.

\section{Analogous to the Landau-type quantization:}

In this section,we start our discussion by describing the quantum dynamics
of a moving electric quadrupole moment interacting with electromagnetic
fields as suggested in \cite{17} . By considering an electric quadrupole
moment as a scalar particle, the potential energy of a multipole expansion
in the rest frame of a particle can be written as

\begin{equation}
U=q\Phi -\overrightarrow{d}.\nabla \Phi +\underset{i,j}{\tsum }%
Q_{ij}\partial _{i}\partial _{j}\Phi
\end{equation}%
where $q$ is the electric charge, $\overrightarrow{d}$ is the electric
dipole moment, $Q_{ij}$ is the electric quadrupole moment tensor, and $\Phi $
is the electric potential.

In order to study the dynamics of an electric quadrupole moment, we consider 
$q=0$ , $\overrightarrow{d}=0$ and $\overrightarrow{E}=-\overrightarrow{%
\nabla }\Phi $, where $\overrightarrow{E}$ is the electric field. Therefore,
the equation (1) reads

\begin{equation}
U=-\underset{i,j}{\tsum }Q_{ij}\partial _{i}E_{j}
\end{equation}%
For a moving quadrupole, Lagrangian of this system (a constant mass system)
becomes

\begin{equation}
L=\frac{1}{2}mv^{2}+\underset{i,j}{\tsum }Q_{ij}\partial _{i}E_{j}
\end{equation}%
Now we must apply the Lorentz transformation of the electromagnetic fields.
Therefore, we replace the electric field in (3) by

\begin{equation}
\overrightarrow{E}\rightarrow \overrightarrow{E}+\frac{1}{c}\overrightarrow{v%
}\times \overrightarrow{B}
\end{equation}%
where $\overrightarrow{E}$ and $\overrightarrow{B}$ are the electric and
magnetic fields, respectively. Thus, Lagrangian (3) becomes

\begin{equation}
L=\frac{1}{2}mv^{2}+\overrightarrow{Q}.\overrightarrow{E}-\frac{1}{c}%
\overrightarrow{v}.(\overrightarrow{Q}\times \overrightarrow{B})
\end{equation}%
where we used

\begin{equation}
Q_{i}=\underset{i,j}{\tsum }Q_{ij}\partial _{j}\quad ,\quad \overrightarrow{Q%
}=\underset{i}{\tsum }Q_{i}\widehat{e}_{i}
\end{equation}%
as in \cite{18,31}. Using the canonical momentum

\begin{equation}
\overrightarrow{P}=m\overrightarrow{v}-\frac{1}{c}(\overrightarrow{Q}\times 
\overrightarrow{B})
\end{equation}%
the classical Hamiltonian of a constant mass reads

\begin{equation}
H=\frac{1}{2m}\left[ \overrightarrow{P}+\frac{1}{c}(\overrightarrow{Q}\times 
\overrightarrow{B})\right] ^{2}-\overrightarrow{Q}.\overrightarrow{E}
\end{equation}

To write the quantum Hamiltonian operator, we replace the canonical momentum 
$\overrightarrow{P}$ by the operator $\widehat{P}=-i\overrightarrow{\nabla }$
for constant mass settings. However, in this work we are interested to study
the PDM system. Thus, we rewrite the PDM-non relativistic Hamiltonian (in $%
\hbar =2m_{\circ }=c=1$ units) as

\begin{equation}
\widehat{H}=\left( \frac{\widehat{P}\left( \overrightarrow{r}\right) +%
\overrightarrow{A}_{eff}\left( \overrightarrow{r}\right) }{\sqrt{m\left( 
\overrightarrow{r}\right) }}\right) ^{2}-\overrightarrow{Q}.\overrightarrow{E%
}
\end{equation}%
where the kinetic energy term was proposed by Mustafa and Algadhi \cite{59}
along with the definition of PDM- momentum operator (which resulted from a
factorization recipe of Mustafa and Mazharimousavi in \cite{46}):

\begin{equation}
\widehat{P}\left( \overrightarrow{r}\right) =-i\left[ \overrightarrow{\nabla 
}-\frac{1}{4}\left( \frac{\overrightarrow{\nabla }m\left( \overrightarrow{r}%
\right) }{m\left( \overrightarrow{r}\right) }\right) \right]
\end{equation}
\ where

\begin{equation}
\overrightarrow{A}_{eff}\left( \overrightarrow{r}\right) =\overrightarrow{Q}%
\times \overrightarrow{B}\quad ,\quad V_{eff}\left( \overrightarrow{r}%
\right) =-\overrightarrow{Q}.\overrightarrow{E}
\end{equation}

In this way, the corresponding time-independent Schr\"{o}dinger equation is
written in the form

\begin{equation}
\left[ \left( \frac{\widehat{P}\left( \overrightarrow{r}\right) +%
\overrightarrow{A}_{eff}\left( \overrightarrow{r}\right) }{\sqrt{m\left( 
\overrightarrow{r}\right) }}\right) ^{2}-\overrightarrow{Q}.\overrightarrow{E%
}\right] \psi \left( \overrightarrow{r}\right) =\varepsilon \psi \left( 
\overrightarrow{r}\right) .
\end{equation}

Hence,

\begin{eqnarray}
&&\left[ \left( \frac{\widehat{P}\left( \overrightarrow{r}\right) }{\sqrt{%
m\left( \overrightarrow{r}\right) }}\right) ^{2}-\frac{2i}{m\left( 
\overrightarrow{r}\right) }\overrightarrow{A}_{eff}\left( \overrightarrow{r}%
\right) \cdot \overrightarrow{\nabla }-\frac{i}{m\left( \overrightarrow{r}%
\right) }\left( \overrightarrow{\nabla }\cdot \overrightarrow{A}_{eff}\left( 
\overrightarrow{r}\right) \right) \right.  \notag \\
&&\left. +i\overrightarrow{A}_{eff}\left( \overrightarrow{r}\right) .\left( 
\frac{\overrightarrow{\nabla }m\left( \overrightarrow{r}\right) }{m\left( 
\overrightarrow{r}\right) ^{2}}\right) +\frac{\overrightarrow{A}_{eff}\left( 
\overrightarrow{r}\right) ^{2}}{m\left( \overrightarrow{r}\right) }-%
\overrightarrow{Q}.\overrightarrow{E}\right] \left. \psi \left( 
\overrightarrow{r}\right) =\varepsilon \psi \left( \overrightarrow{r}\right)
\right.
\end{eqnarray}%
in which the vector potential satisfies the Coulomb gauge $\overrightarrow{%
\nabla }\cdot \overrightarrow{A}_{eff}=0$. Moreover, using the momentum
operator in equation(10) would imply:

\begin{eqnarray}
&&\left[ -\frac{1}{m\left( \overrightarrow{r}\right) }\overrightarrow{\nabla 
}^{2}+\left( \frac{\overrightarrow{\nabla }m\left( \overrightarrow{r}\right) 
}{m\left( \overrightarrow{r}\right) ^{2}}\right) \cdot \overrightarrow{%
\nabla }+\frac{1}{4}\left( \frac{\overrightarrow{\nabla }^{2}m\left( 
\overrightarrow{r}\right) }{m\left( \overrightarrow{r}\right) ^{2}}\right) -%
\frac{7}{16}\left( \frac{\left[ \overrightarrow{\nabla }m\left( 
\overrightarrow{r}\right) \right] ^{2}}{m\left( \overrightarrow{r}\right)
^{3}}\right) -\frac{2\ i\ }{m\left( \overrightarrow{r}\right) }%
\overrightarrow{A}_{eff}\left( \overrightarrow{r}\right) \cdot 
\overrightarrow{\nabla }\right.  \notag \\
&&\qquad \quad \qquad \qquad \left. +i\ \ \overrightarrow{A}_{eff}\left( 
\overrightarrow{r}\right) .\left( \frac{\overrightarrow{\nabla }m\left( 
\overrightarrow{r}\right) }{m\left( \overrightarrow{r}\right) ^{2}}\right) +%
\frac{\overrightarrow{A}_{eff}\left( \overrightarrow{r}\right) ^{2}}{m\left( 
\overrightarrow{r}\right) }-\overrightarrow{Q}.\overrightarrow{E}\right]
\left. \psi \left( \overrightarrow{r}\right) =\varepsilon \psi \left( 
\overrightarrow{r}\right) .\right.
\end{eqnarray}

The discussion of the possibility of achieving the Landau quantization for
an electric quadrupole moment was done in \cite{18}, where they found that
the Landau quantization can be achieved by imposing these two conditions:
the first one is that the tensor $Q_{ij}$ must be symmetric and tracless.
And the second one is that the field configuration must be chosen in such a
way that there exists a uniform effective magnetic field given by

\begin{equation}
\overrightarrow{B}_{eff}=\overrightarrow{\nabla }\times \overrightarrow{A}%
_{eff}=\overrightarrow{\nabla }\times (\overrightarrow{Q}\times 
\overrightarrow{B})=constant\ vector
\end{equation}%
perpendicular to the plane of motion of the electric quadrupole moment.
Thus, it is clear that the field configuration depends on the choice of the
components of the tensor $Q_{ij}$ that describes the electric quadrupole
moment. Moreover, $\overrightarrow{E}$ must satisfy the electrostatic
conditions $\left( \overrightarrow{\nabla }\times \overrightarrow{E}%
=0~,~\partial _{t}\overrightarrow{E}=0\right) .$

In the following, we present the field configurations and structures of the
tensor $Q_{ij}$. Thus, we choose the case when the tensor $Q_{ij}$ has the
non-null components:

\begin{equation}
Q_{\rho \rho }=Q_{\phi \phi }=Q,~~Q_{zz}=-2Q\quad (diagonal~form)
\end{equation}%
which was studied by \cite{17,18}, where $Q$ is a constant. It is notable
that this choice satisfies the properties of the tensor $Q_{ij}.$ Moreover,
we consider a magnetic field given by \cite{18,31}

\begin{equation}
\overrightarrow{B}=\frac{1}{2}B_{\circ }\rho ^{2}\ \widehat{z}
\end{equation}%
where $B_{\circ }$ is a constant. By using the the definitions of (6) and
the assumption in (16) we obtain the electric quadrupole moment as

\begin{equation}
\overrightarrow{Q}=\left( Q\partial _{\rho }\right) \widehat{\rho }\ +\left(
Q\partial _{\phi }\right) \widehat{\phi }-\left( 2Q\partial _{z}\right) 
\widehat{z}
\end{equation}%
At this point, we can find the effective vector potential $\overrightarrow{A}%
_{eff}$ as

\begin{equation}
\overrightarrow{A}_{eff}\left( \overrightarrow{r}\right) =\overrightarrow{Q}%
\times \overrightarrow{B}=-QB_{\circ }\rho \widehat{\phi }
\end{equation}%
Consequently, the effective magnetic field reads

\begin{equation}
\overrightarrow{B}_{eff}\left( \overrightarrow{r}\right) =\overrightarrow{%
\nabla }\times \overrightarrow{A}_{eff}\left( \overrightarrow{r}\right)
=-2QB_{\circ }\widehat{z}
\end{equation}%
which satisfies the second condition (15), where $\overrightarrow{B}_{eff}$
is a uniform effective magnetic field.

We may now discuss the separability of the PDM Schr\"{o}dinger equation(14)
in the cylindrical coordinates $\left( \rho ,\phi ,z\right) $ and under
azimuthal symmetrization. By assuming that the field configurations and that
the PDM functions are only radially dependent \cite{54,55,59,60,61} (i,e., $%
m\left( \overrightarrow{r}\right) =M\left( \rho ,\phi ,z\right) =g\left(
\rho \right) $ ), the wavefunction can be written as

\begin{equation}
\psi \left( \rho ,\phi ,z\right) =e^{im\phi }e^{ikz}R\left( \rho \right)
\end{equation}%
where $m=0,\pm 1,\pm 2,....,\pm \ell $ is the magnetic quantum number.
Thereby, and with the substitution of (19),(20) and (21) into (14), we
obtain the radial equation:

\begin{eqnarray}
&&\left. \frac{R^{\prime \prime }\left( \rho \right) }{R\left( \rho \right) }%
-\left( \frac{g^{\prime }\left( \rho \right) }{g\left( \rho \right) }-\frac{1%
}{\rho }\right) \frac{R^{\prime }\left( \rho \right) }{R\left( \rho \right) }%
-\frac{1}{4}\left( \frac{g^{\prime \prime }\left( \rho \right) }{g\left(
\rho \right) }-\frac{g^{\prime }\left( \rho \right) }{\rho g\left( \rho
\right) }\right) +\frac{7}{16}\left( \frac{g^{\prime }\left( \rho \right) }{%
g\left( \rho \right) }\right) ^{2}\right.  \notag \\
&&\qquad \qquad \qquad \left. +g\left( \rho \right) (\varepsilon +%
\overrightarrow{Q}.\overrightarrow{E})-\frac{m^{2}}{\rho ^{2}}-Q^{2}B_{\circ
}^{2}\rho ^{2}+2QB_{\circ }m-k_{z}^{2}=0.\right.
\end{eqnarray}%
Which is to be solved for no electric field $\overrightarrow{E}=0$ and for a
different choice of radial electric fields $\overrightarrow{E}\neq 0,$ with
suitable PDM settings, to find the exact eigenvalues and eigenfunctions of
the system.

\section{\protect\bigskip PDM particles with an electric quadrupole moment
in a magnetic field:}

In this section,we focus on the discussion of Landau quantization for an
electric quadrupole moment interacting with an external magnetic field, and
a vanishing electric field $\overrightarrow{E}=0$ $\left(
i.e.~~V_{eff}=0\right) .$ At this point, equation(22) would read

\begin{eqnarray}
&&R^{\prime \prime }\left( \rho \right) +\left[ -\left( \frac{g^{\prime
}\left( \rho \right) }{g\left( \rho \right) }-\frac{1}{\rho }\right)
R^{\prime }\left( \rho \right) -\frac{1}{4}\left( \frac{g^{\prime \prime
}\left( \rho \right) }{g\left( \rho \right) }-\frac{g^{\prime }\left( \rho
\right) }{\rho g\left( \rho \right) }\right) +\frac{7}{16}\left( \frac{%
g^{\prime }\left( \rho \right) }{g\left( \rho \right) }\right) ^{2}\right. 
\notag \\
&&\qquad \qquad \qquad \left. +g\left( \rho \right) \varepsilon -\frac{m^{2}%
}{\rho ^{2}}-Q^{2}B_{\circ }^{2}\rho ^{2}+2QB_{\circ }m-k_{z}^{2}\right]
R\left( \rho \right) =0.
\end{eqnarray}%
In the following examples, we use some power-low PDM type in the radial Schr%
\"{o}dinger equation (23) and report their exact-solutions.

\subsection{ \protect\bigskip Model-I: A linear-type PDM $g\left( \protect%
\rho \right) =\protect\eta \protect\rho $ :}

Consider a neutral particle with the radial cylindrical PDM setting, $%
g\left( \rho \right) =\eta \rho ,$ and the electric quadrupole moment of
(18) in presence of the magnetic field in (17). Then, the Schr\"{o}dinger
equation(23) would read

\begin{equation}
R^{\prime \prime }\left( \rho \right) +\left[ \frac{-\left(
m^{2}-3/16\right) }{\rho ^{2}}-Q^{2}B_{\circ }^{2}\rho ^{2}+\eta \rho
\varepsilon +2QB_{\circ }m-k_{z}^{2}\right] R\left( \rho \right) =0
\end{equation}%
Now, let us make a simple change of variables in equation (24) and use $r=%
\sqrt{QB_{\circ }}\rho $. Then equation (24) becomes

\begin{equation}
R^{\prime \prime }\left( r\right) +\left[ \frac{-\left( m^{2}-3/16\right) }{%
r^{2}}-r^{2}+\frac{\eta \varepsilon }{\left( QB_{\circ }\right) ^{3/2}}r+%
\frac{2QB_{\circ }m-k_{z}^{2}}{QB_{\circ }}\right] R\left( r\right) =0,
\end{equation}%
which implies the one-dimensional Schr\"{o}dinger form of the Biconfluent
Heun equation ( see, \cite{63,64}) reads

\smallskip

\begin{equation}
R^{\prime \prime }\left( r\right) +\left[ \frac{\left( 1-\alpha ^{2}\right) 
}{4r^{2}}-\frac{1}{2r}\delta -\beta r-r^{2}+\gamma -\frac{\beta ^{2}}{4}%
\right] R\left( r\right) =0
\end{equation}%
where

\begin{equation}
\frac{1}{4}\left( 1-\alpha ^{2}\right) =3/16-m^{2},~\beta =\frac{-\eta
\varepsilon }{\left( QB_{\circ }\right) ^{3/2}},~\frac{\delta }{2}=0,\quad
\gamma -\frac{\beta ^{2}}{4}=\frac{2QB_{\circ }m-k_{z}^{2}}{QB_{\circ }}
\end{equation}

To solve the above equation $\left( 26\right) $, we consider the asymptotic
behavior for $r\rightarrow 0~and~r\rightarrow \infty ,$ the function $%
R\left( r\right) $ can be written in terms of an unknown function $u\left(
r\right) $ as follows

\begin{equation}
R\left( r\right) =r^{\left( 1+\alpha \right) /2}e^{-\left( \beta
r+r^{2}\right) /2}u\left( r\right)
\end{equation}%
that transforms equation$\left( 26\right) $ into a simpler form

\begin{equation}
ru^{\prime \prime }\left( r\right) +\left[ 1+\alpha -\beta r-2r^{2}\right]
u^{\prime }\left( r\right) +\left[ \left( \gamma -2-\alpha \right) r-\frac{1%
}{2}\left( \delta +\left( 1+\alpha \right) \beta \right) \right] u\left(
r\right) =0
\end{equation}%
which is the Biconfluent Heun-type equation (BHE) \cite{64}, where $\alpha
,\beta ,\gamma $ and $\delta $ are arbitrary parameters. The polynomial
solutions of this equation (c.f, e.g., \cite{62,63,64,65,66}) is

\begin{equation}
u\left( r\right) =H_{B}\left( \alpha ,\beta ,\gamma ,\delta ;r\right)
\end{equation}%
where $H_{B}\left( \alpha ,\beta ,\gamma ,\delta ;r\right) $ are the Heun
polynomials of degree $n_{\rho }$ such that

\begin{equation}
\gamma -2-\alpha =2n_{\rho },\quad where\quad n_{\rho }=0,1,2,...,\quad
and\quad a_{n_{\rho }+1}=0~.
\end{equation}%
Here, $n_{\rho }$ is the radial quantum number and $a_{n_{\rho }+1}$ is a
polynomial of degree $n_{\rho }+1$ defined by the recurrent relation (see 
\cite{64,65,66} for more details). By substituting $\left( 27\right) $ into $%
\left( 31\right) $, we get the exact eigenvalues

\begin{equation}
\varepsilon _{n_{\rho },m}=\frac{\left( 2QB_{\circ }\right) ^{3/2}}{\eta }%
\left[ 1+n_{\rho }-m+\sqrt{m^{2}+1/16}+\frac{k_{z}^{2}}{2QB_{\circ }}\right]
^{1/2}
\end{equation}%
where the cyclotron frequency is $\omega =\frac{\left( 2QB_{\circ }\right)
^{3/2}}{\eta },$ and the exact normalized eigenfunctuons is

\begin{equation}
R\left( \rho \right) =N\rho ^{\left\vert \widetilde{\ell }\right\vert
+1/2}e^{-\left( \frac{Q^{2}B_{\circ }^{2}\rho ^{2}-\eta \varepsilon \rho }{%
2QB_{\circ }}\right) }H_{B}\left( \alpha ,\beta ,\gamma ,0;\sqrt{QB_{\circ }}%
\rho \right)
\end{equation}%
where $N$ is the normalization constant, $\left\vert \widetilde{\ell }%
\right\vert =\sqrt{m^{2}+1/16}$, and $\alpha ,\beta ~and~\gamma $ are
defined respectively in (27).

Comparing with \cite{31}, the eigenvalues are changed due to the effect of
the PDM of the system, \ where the spectrum of energy (32) is proportional
to $n^{1/2}$ and removes degeneracies (associated with the magnetic quantum
number) similar to the energy levels reported in \cite{31}, where they are
proportional to $n$. Furthermore, the frequency is also modified.

\subsection{\protect\bigskip Model-II: A harmonic-type PDM $g\left( \protect%
\rho \right) =\protect\eta \protect\rho ^{2}$:}

A PDM neutral particle with $g\left( \rho \right) =\eta \rho ^{2},$ and an
electric quadrupole moment interacting with the external magnetic field (17)
would imply that equation (23) be rewritten as

\begin{equation}
R^{\prime \prime }\left( \rho \right) -\frac{1}{\rho }R^{\prime }\left( \rho
\right) +\left[ \frac{-\left( m^{2}-3/4\right) }{\rho ^{2}}-\left(
Q^{2}B_{\circ }^{2}-\eta \varepsilon \right) \rho ^{2}+2QB_{\circ
}m-k_{z}^{2}\right] R\left( \rho \right) =0
\end{equation}%
To determine the radial part $R\left( \rho \right) $ of the wave function
and the energy spectrum, we follow the same analysis of Gasiorowicz \cite{68}
and start by using the change of variable $x=\left( Q^{2}B_{\circ }^{2}-\eta
\varepsilon \right) ^{1/4}\rho $ in (34) to obtain

\begin{equation}
R^{\prime \prime }\left( x\right) -\frac{1}{x}R^{\prime }\left( x\right) -%
\frac{L^{2}}{x^{2}}R\left( x\right) -x^{2}R\left( x\right) +\mu R\left(
x\right) =0
\end{equation}%
where

\begin{equation}
L^{2}=m^{2}-3/4\quad and\quad \mu =\frac{2QB_{\circ }m-k_{z}^{2}}{\left(
Q^{2}B_{\circ }^{2}-\eta \varepsilon \right) ^{1/2}}
\end{equation}%
Next, we consider the asymptotic solutions $\left( x\rightarrow
0~and~x\rightarrow \infty \right) $ of the radial wavefunction $R\left(
x\right) $ to come out with

\begin{equation}
R\left( x\right) =x^{1+\left\vert \widetilde{\ell }\right\vert
}e^{-x/2}G\left( x\right)
\end{equation}%
with

\begin{equation}
\widetilde{\ell }^{2}=L^{2}+1\Longrightarrow ~\left\vert \widetilde{\ell }%
\right\vert =\sqrt{m^{2}+1/4},\quad with\ \ \widetilde{\ell }>0.
\end{equation}%
Substituting (37) in (35) would imply

\begin{equation}
G^{\prime \prime }\left( x\right) +\left( \frac{1+2\left\vert \widetilde{%
\ell }\right\vert }{x}-2x\right) G^{\prime }\left( x\right) +\left( \mu
-2-2\left\vert \widetilde{\ell }\right\vert \right) G\left( x\right) =0
\end{equation}%
Which, in turn, with $y=x^{2}$ yields

\begin{equation}
yG^{\prime \prime }\left( y\right) +\left( 1+\left\vert \widetilde{\ell }%
\right\vert -y\right) G^{\prime }\left( y\right) +\left( \frac{\mu }{4}-%
\frac{\left\vert \widetilde{\ell }\right\vert }{2}-\frac{1}{2}\right)
G\left( y\right) =0
\end{equation}%
this equation is the confluent hypergeometric equation, the series of which
is a polynomial of degree $n_{\rho }$ (finite everywhere) when

\begin{equation}
n_{\rho }=\frac{\mu }{4}-\frac{\left\vert \widetilde{\ell }\right\vert }{2}-%
\frac{1}{2}~.
\end{equation}%
Consequently, (36) and (38) would give the eigenvalues as

\begin{equation}
\varepsilon _{n_{\rho },m}=\frac{Q^{2}B_{\circ }^{2}}{\eta }\left[ 1-\left( 
\frac{\frac{k_{z}^{2}}{2QB_{\circ }}-m}{1+2n_{\rho }+\sqrt{m^{2}+1/4}}%
\right) ^{2}\right]
\end{equation}%
and the eigenfunctions as

\begin{equation}
R\left( \rho \right) =N\rho ^{\left\vert \widetilde{\ell }\right\vert +1}e^{-%
\frac{\sqrt{Q^{2}B_{\circ }^{2}-\eta \varepsilon }}{2}\rho
^{2}}{}_{1}F_{1}\left( -n_{\rho };\left\vert \widetilde{\ell }\right\vert +1;%
\sqrt{Q^{2}B_{\circ }^{2}-\eta \varepsilon }\rho ^{2}\right) .
\end{equation}

In this case, the effect of PDM setting produces a new contribution to the
non-degenerate energy levels (42), where they are proportional to $n^{-2}$
and the frequency is modified as $\varpi =\frac{Q^{2}B_{\circ }^{2}}{\eta }.$

\section{PDM-particles with an electric quadrupole moment and
electromagnetic fields:}

In this section,we study PDM-particles with an electric quadrupole moment
and electromagnetic fields. However, we focus on analysis of the effect of
the electric field on a PDM particle with an electric quadrupole moment in
the presence of a magnetic field (on the Landau-type system reported in the
previous section). For this purpose, we choose a sample of radial electric
fields (c.f., e.g., \cite{14,15,16,24,31}), in the following illustrative
examples.

\subsection{\protect\bigskip The influence of A Coulomb-type electric field
on the Landau-type system:}

Consider a radial electric field in the form of

\begin{equation}
\overrightarrow{E}=\frac{\lambda }{\rho }\widehat{\rho }
\end{equation}%
where $\lambda $ is a constant \cite{31}.

Thus, we can see that the interaction between the electric quadrupole moment
(18) and the electric field (44) leads to an effective scalar potential

\begin{equation}
V_{eff}\left( \rho \right) =-\overrightarrow{Q}.\overrightarrow{E}=\frac{%
Q\lambda }{\rho ^{2}}
\end{equation}%
which plays the role of a scalar potential in the PDM-Schr\"{o}dinger
equation (22) to imply

\begin{eqnarray}
&&R^{\prime \prime }\left( \rho \right) -\left( \frac{g^{\prime }\left( \rho
\right) }{g\left( \rho \right) }-\frac{1}{\rho }\right) R^{\prime }\left(
\rho \right) +\left[ -\frac{1}{4}\left( \frac{g^{\prime \prime }\left( \rho
\right) }{g\left( \rho \right) }-\frac{g^{\prime }\left( \rho \right) }{\rho
g\left( \rho \right) }\right) +\frac{7}{16}\left( \frac{g^{\prime }\left(
\rho \right) }{g\left( \rho \right) }\right) ^{2}\right.  \notag \\
&&\qquad ~~+\left. g\left( \rho \right) \varepsilon -g\left( \rho \right) 
\frac{Q\lambda }{\rho ^{2}}-\frac{m^{2}}{\rho ^{2}}-Q^{2}B_{\circ }^{2}\rho
^{2}+2QB_{\circ }m-k_{z}^{2}\right] R\left( \rho \right) =0.
\end{eqnarray}

Hereby, we again consider the same examples used in the previous section to
find exact solutions of equation (46):

\subsubsection{ Model-I : $g\left( \protect\rho \right) =\protect\eta 
\protect\rho $ :}

The PDM radial Schr\"{o}dinger equation (46) with, $g\left( \rho \right)
=\eta \rho ,$ reads

\ \ 
\begin{equation}
R^{\prime \prime }\left( \rho \right) +\left[ \frac{-\left(
m^{2}-3/16\right) }{\rho ^{2}}-Q^{2}B_{\circ }^{2}\rho ^{2}+\eta \rho
\varepsilon -\frac{\eta \lambda Q}{\rho }+2QB_{\circ }m-k_{z}^{2}\right]
R\left( \rho \right) =0
\end{equation}%
with the change of variable $r=\sqrt{QB_{\circ }}\rho ,$ equation (47)
becomes

\begin{equation}
R^{\prime \prime }\left( r\right) +\left[ \frac{-\left( m^{2}-3/16\right) }{%
r^{2}}-r^{2}+\frac{\eta \varepsilon }{\left( QB_{\circ }\right) ^{3/2}}r-%
\frac{\eta \lambda Q}{\left( QB_{\circ }\right) ^{1/2}r}+\frac{2QB_{\circ
}m-k_{z}^{2}}{QB_{\circ }}\right] R\left( r\right) =0
\end{equation}%
To find its solutions, we define these parameters

\begin{equation}
\frac{1}{4}\left( 1-\alpha ^{2}\right) =3/16-m^{2},~\beta =\frac{-\eta
\varepsilon }{\left( QB_{\circ }\right) ^{3/2}},~\frac{\delta }{2}=\frac{%
\eta \lambda Q}{\left( QB_{\circ }\right) ^{1/2}},\quad \gamma -\frac{\beta
^{2}}{4}=\frac{2QB_{\circ }m-k_{z}^{2}}{QB_{\circ }},
\end{equation}%
and follow the same steps as in (28) to (31). Thus the exact eigenvalues are

\begin{equation}
\varepsilon _{n_{\rho },m}=\frac{\left( 2QB_{\circ }\right) ^{3/2}}{\eta }%
\left[ \left( 1+n_{\rho }+-m+\sqrt{m^{2}+1/16}\right) +\frac{k_{z}^{2}}{%
2QB_{\circ }}\right] ^{1/2}
\end{equation}%
and the exact eigenfunctions are

\begin{equation}
R\left( \rho \right) =N\rho ^{\left\vert \widetilde{\ell }\right\vert
+1/2}e^{-\left( \frac{Q^{2}B_{\circ }^{2}\rho ^{2}-\eta \varepsilon \rho }{%
2QB_{\circ }}\right) }H_{B}\left( \alpha ,\beta ,\gamma ,\delta ;\sqrt{%
QB_{\circ }}\rho \right)
\end{equation}

\smallskip It is obvious that these eigenvalues (50) are the same as the
eigenvalues in the absence of an electric field for the same PDM setting
given in (32) but with different eigenfunctions. Thus, the effective
potential with the PDM setting does not effect the eigenvalues of the system.

\subsubsection{Model-II: $g\left( \protect\rho \right) =\protect\eta \protect%
\rho ^{2}$ :}

The substitution of $g\left( \rho \right) =\eta \rho ^{2},$ in the PDM
radial Schr\"{o}dinger equation (46) would yield

\begin{equation}
R^{\prime \prime }\left( \rho \right) -\frac{1}{\rho }R^{\prime }\left( \rho
\right) +\left[ \frac{-\left( m^{2}-3/4\right) }{\rho ^{2}}-\left(
Q^{2}B_{\circ }^{2}-\eta \varepsilon \right) \rho ^{2}-\eta \lambda
Q+2QB_{\circ }m-k_{z}^{2}\right] R\left( \rho \right) =0
\end{equation}%
We repeat the same procedure as in the previous section and immediately
write the corresponding eigenvalues and radial wave functions, respectively,
as

\begin{equation}
\varepsilon _{n_{\rho },m}=\frac{Q^{2}B_{\circ }^{2}}{\eta }\left[ 1-\left( 
\frac{\frac{\eta \lambda Q+k_{z}^{2}}{2QB_{\circ }}-m}{1+2n_{\rho }+\sqrt{%
m^{2}+1/4}}\right) ^{2}\right]
\end{equation}%
and

\begin{equation}
R\left( \rho \right) =N\rho ^{\left\vert \widetilde{\ell }\right\vert +1}e^{-%
\frac{\sqrt{Q^{2}B_{\circ }^{2}-\eta \varepsilon }}{2}\rho
^{2}}{}_{1}F_{1}\left( -n_{\rho };\left\vert \widetilde{\ell }\right\vert +1;%
\sqrt{Q^{2}B_{\circ }^{2}-\eta \varepsilon }\rho ^{2}\right)
\end{equation}%
where $\left\vert \widetilde{\ell }\right\vert =\sqrt{m^{2}+1/4},$ with$\ \ 
\widetilde{\ell }>0.$ In this case, the influence of the PD-effective
potential is appeared by making a shift in the energy levels of (42) and
producing new eigenvalues (53), therefore.

\subsection{The influence of A linear-type electric field on the Landau-type
system:}

Now, let us consider another radial electric field (e.g., \cite{14,15,16,24}%
) as

\begin{equation}
\overrightarrow{E}=\frac{\lambda \rho }{2}~\widehat{\rho }
\end{equation}%
Using the same components of the the electric quadrupole moment tensor
defined in (16), the effective scalar potential given in the PDM-Schr\"{o}%
dinger equation(22) becomes

\begin{equation}
V_{eff}\left( \rho \right) =-\overrightarrow{Q}.\overrightarrow{E}=-\frac{%
Q\lambda }{2}
\end{equation}%
Hence, equation(22) reads

\begin{eqnarray}
&&R^{\prime \prime }\left( \rho \right) -\left( \frac{g^{\prime }\left( \rho
\right) }{g\left( \rho \right) }-\frac{1}{\rho }\right) R^{\prime }\left(
\rho \right) +\left[ -\frac{1}{4}\left( \frac{g^{\prime \prime }\left( \rho
\right) }{g\left( \rho \right) }-\frac{g^{\prime }\left( \rho \right) }{\rho
g\left( \rho \right) }\right) +\frac{7}{16}\left( \frac{g^{\prime }\left(
\rho \right) }{g\left( \rho \right) }\right) ^{2}\right.  \notag \\
&&\qquad ~~+\left. g\left( \rho \right) \varepsilon +g\left( \rho \right) 
\frac{Q\lambda }{2}-\frac{m^{2}}{\rho ^{2}}-Q^{2}B_{\circ }^{2}\rho
^{2}+2QB_{\circ }m-k_{z}^{2}\right] R\left( \rho \right) =0.
\end{eqnarray}

In the two examples below, we investigate the influence of this effective
potential using the same PDM setting that used in the previous sections:

\subsubsection{\protect\bigskip Model-I: $g\left( \protect\rho \right) =%
\protect\eta \protect\rho :$}

With $g\left( \rho \right) =\eta \rho $ in (57), we obtain

\ \ 
\begin{equation}
R^{\prime \prime }\left( \rho \right) +\left[ \frac{-\left(
m^{2}-3/16\right) }{\rho ^{2}}-Q^{2}B_{\circ }^{2}\rho ^{2}+\left( \eta
\varepsilon +\frac{\eta \lambda Q}{2}\right) \rho +2QB_{\circ }m-k_{z}^{2}%
\right] R\left( \rho \right) =0
\end{equation}

\smallskip Using the previous technique for the linear-type PDM to get the
exact solutions for this case. Hence, this would correspond to the exact
eigenvalues and eigenfunctions given, respectively, 

\begin{equation}
\varepsilon _{n_{\rho },m}=\frac{\left( 2QB_{\circ }\right) ^{3/2}}{\eta }%
\left[ 1+n_{\rho }-m+\sqrt{m^{2}+1/16}+\frac{k_{z}^{2}}{2QB_{\circ }}\right]
^{1/2}-\frac{\lambda Q}{2}
\end{equation}

\begin{equation}
R\left( \rho \right) =N\rho ^{\left\vert \widetilde{\ell }\right\vert
+1/2}e^{-\frac{Q^{2}B_{\circ }^{2}\rho ^{2}-\left( \eta \varepsilon +\frac{%
\eta \lambda Q}{2}\right) \rho }{2QB_{\circ }}}H_{B}\left( \alpha ,\beta
,\gamma ,0;\sqrt{QB_{\circ }}\rho \right) 
\end{equation}%
where, $\left\vert \widetilde{\ell }\right\vert =\sqrt{m^{2}+1/16}$ and the
parameters $\alpha ,\beta ,\gamma ~and~\delta $ are defined as

\begin{equation}
\frac{1}{4}\left( 1-\alpha ^{2}\right) =3/16-m^{2},\ \ \beta =\frac{-\left(
\eta \varepsilon +\frac{\eta \lambda Q}{2}\right) }{\left( QB_{\circ
}\right) ^{3/2}},\ \ \gamma -\frac{\beta ^{2}}{4}=\frac{2QB_{\circ
}m-k_{z}^{2}}{QB_{\circ }},\ \ \frac{\delta }{2}=0.
\end{equation}

\subsubsection{Model-II: $g\left( \protect\rho \right) =\protect\eta \protect%
\rho ^{2}$ :}

Considering this radial cylindrical PDM would imply that equation (57) be
rewritten as

\begin{equation}
R^{\prime \prime }\left( \rho \right) -\frac{1}{\rho }R^{\prime }\left( \rho
\right) +\left[ \frac{-\left( m^{2}-3/4\right) }{\rho ^{2}}-\left(
Q^{2}B_{\circ }^{2}-\frac{\eta \lambda Q}{2}-\eta \varepsilon \right) \rho
^{2}+2QB_{\circ }m-k_{z}^{2}\right] R\left( \rho \right) =0
\end{equation}

Equation (62) is again in the same form of equation (34) and admits the
exact solution of the eigenvalues and the corresponding radial
eigenfunctions as

\begin{equation}
\varepsilon _{n_{\rho },m}=\frac{Q^{2}B_{\circ }^{2}}{\eta }\left[ 1-\left( 
\frac{\frac{k_{z}^{2}}{2QB_{\circ }}-m}{1+2n_{\rho }+\sqrt{m^{2}+1/4}}%
\right) ^{2}\right] -\frac{\lambda Q}{2}
\end{equation}%
and

\begin{equation}
R\left( \rho \right) =N\rho ^{\widetilde{\left\vert \ell \right\vert }+1}e^{-%
\frac{\sqrt{Q^{2}B_{\circ }^{2}-\frac{\eta \lambda Q}{2}-\eta \varepsilon }}{%
2}\rho ^{2}}{}_{1}F_{1}\left( -n_{\rho };\left\vert \widetilde{\ell }%
\right\vert +1;\sqrt{Q^{2}B_{\circ }^{2}-\frac{\eta \lambda Q}{2}-\eta
\varepsilon }\rho ^{2}\right)
\end{equation}%
where $\left\vert \widetilde{\ell }\right\vert =\sqrt{m^{2}+1/4},$ with$\ \ 
\widetilde{\ell }>0.$

It is shown that the effective potential generated by the interaction
between the electric quadrupole moment and the radial electric field given
by (55) produces constant potential ($-\frac{Q\lambda }{2}$). Thus, the
effect of mass settings on the effective potential in the term ($g\left(
\rho \right) \overrightarrow{Q}.\overrightarrow{E}$) yields only a constant
shift in the energy levels given in (32) and (42) creating\ a new set of
energies given in (59) and (63).

\section{Concluding Remarks}

In this paper, we have started with a quantum system of an electric
quadrupole moment interacting with magnetic and electric fields of a
constant mass, as has been previously reported in the literature \cite{18,31}%
. Next, we have extended this procedure to study PDM systems by using recent
results of \cite{58,59} for the PDM- minimal-coupling and the PDM- momentum
operator given by (9) and (10), respectively. We have discussed the
possibility of achieving the Landau quantization following \cite{18} and
modified this discussion to include a PDM case. Thus, we have recollected
the most important and essential relations ( equations (15)-(20) above),
that have been reported in \cite{18}. Moreover, we have studied this problem
within the context of cylindrical coordinates and investigated the exact
solvability of the PDM radial Schr\"{o}dinger equation of a neutral particle
possessing an electric quadrupole moment interacting with external fields,
where we have considered PDM settings $m\left( \overrightarrow{r}\right)
=g\left( \rho \right) $, along with the field configurations ( documented in
(17) for the magnetic field $\overrightarrow{B}\left( \rho \right) $ and
(44),(55) for electric fields $\overrightarrow{E}\left( \rho \right) $),
which exhibits a pure radial cylindrical dependence. We have shown that the
Landau quantization is produced from the interaction between the magnetic
field and the electric quadrupole moment given in (17) and (18)
respectively, where this has yielded the PDM radial Schr\"{o}dinger equation
of this system in the absence of an electric field (documented in (23)
above). However, comparing with \cite{18}, the energy levels of the Landau
quantization are modified because of the influence of the spatial dependence
of the mass, where we have obtained different eigenvalues for the two
examples of PDM settings (i.e. $g\left( \rho \right) =\eta \rho $ and $%
g\left( \rho \right) =\eta \rho ^{2})$ given in (32) and (42), respectively.

Furthermore, we have analyzed the effect of the interaction of radial
electric fields with the electric quadrupole moment of a PDM neutral
particle by choosing two particular cases of the electric field $(%
\overrightarrow{E}=\frac{\lambda }{\rho }\widehat{\rho }\ and\ 
\overrightarrow{E}=\frac{\lambda \rho }{2}\widehat{\rho }),$ which have
produced the effective potentials. It is shown that the structure of the PD
effective potential term, plays the role of scalar potentials in the PDM
radial Schr\"{o}dinger equation (22), producing same or an energy shift in
energy levels of the systems with the same mass settings and in the absence
of the effective potential. We have observed that the difference in energy
levels depends on the mass structure. The more complex the chosen mass, the
more this difference will be over Landau levels. However, the exact
eigenvalues and eigenfunctions for all these cases are obtained.

Finally, although the presence of an effective uniform magnetic field
produces the Landau quantization, the influence of the spatial dependence of
the mass of the system yields a new contribution to energy levels creating a
set of new eigenvalues.

This study has investigated, for the first time, that the PDM quantum
particle that possesses multipole moments under the influence of external
fields. Thus, this work opens new discussions regarding the
position-dependent concept and provides a good starting point for future
research.

\end{document}